\documentclass[manuscript, noblind]{geophysics}

\usepackage{amsmath}
\usepackage{url}
\begin{document}

\title{Efficient Upside-Down Rayleigh-Marchenko Imaging through Self-Supervised Focusing Function Estimation}

\renewcommand{\thefootnote}{\fnsymbol{footnote}} 

\address{
\textsuperscript{1}Division of Earth Science and Engineering (ErSE), 
King Abdullah University of Science and Technology (KAUST),
23955-6900, Thuwal, Saudi Arabia, 
{ning.wang.2, tariq.alkhalifah}@kaust.edu.sa \\
\textsuperscript{2}Shearwater GeoServices, RH6 0PA, Gatwick, United Kingdom, mravasi@shearwatergeo.com 
}
\author{
N. Wang\textsuperscript{1}, 
M. Ravasi\textsuperscript{2,1}, 
T. Alkhalifah\textsuperscript{1}
}

\footer{Example}
\lefthead{Wang, Ravasi \& Alkhalifah}
\righthead{Efficient UD-RM Imaging}


\begin{abstract}
The Upside-Down Rayleigh-Marchenko (UD-RM) method has recently emerged as a powerful tool for retrieving subsurface wavefields and images free from artifacts caused by both internal and surface-related multiples. Its ability to handle acquisition setups with large cable spacing or sparse node geometries makes it particularly suitable for ocean-bottom seismic data processing. However, the widespread application of the method is limited by the high computational cost required to estimate the focusing functions, especially when dealing with large imaging domains. To address this limitation, a self-supervised learning approach is proposed to accelerate the estimation of the focusing functions. Specifically, a U-Net network is trained on a small subset of randomly selected image points from within the target area of interest, whose focusing functions are pre-computed using the conventional iterative scheme. The network is tasked to predict both the up- and down-going focusing functions from an initial estimate of the subsurface wavefields. Once trained, the network generalizes to remaining unseen imaging locations, enabling direct prediction of the focusing functions. Validation on a synthetic dataset with both dense and sparse receiver sampling using progressively fewer training points demonstrates the method's effectiveness. In both cases, the resulting images closely match those obtained from the UD-RM method with focusing functions retrieved by the conventional iterative approach at a much lower cost and significantly outperform mirror migration (when the same input dataset is used). Finally, an application to the Volve field data confirms the method’s robustness in practical scenarios. The proposed approach enables seismic imaging at a fraction of the computational cost of the conventional UD-RM approach while maintaining imaging quality, underscoring its potential for large-scale seismic applications.
\end{abstract}

\section{Introduction}
The Marchenko redatuming method~\citep{Broggini2012, Wapenaar2014} enables the retrieval of surface-to-subsurface Green’s functions that contain both primary reflections and internal multiples. By leveraging surface reflection data and a smooth migration velocity model, the method effectively suppresses artifacts associated with incorrect handling of internal multiples during imaging. This is accomplished by solving the Marchenko equations for the so-called focusing functions, which establish a connection between the Green’s functions (from the subsurface focal points to the surface receivers) and surface recorded seismic data. While the Marchenko method has been successfully applied to numerous field datasets~\citep{Ravasi2015, Ravasi2016, Jia2017, Staring2021}, its practical application remains computationally intensive, as focusing functions must be computed independently for each imaging point by evaluating a truncated Neumann series~\citep{vanderNeut2015} or via an iterative scheme such as least squares with QR factorization (LSQR)~\citep{Ravasi2021}. Building on this foundation, the Rayleigh-Marchenko (RM) method~\citep{Ravasi2017} has been proposed to overcome several limitations of the original Marchenko scheme. It allows sources and receivers to be placed at different depths, accounts for the presence of surface-related multiples, and eliminates the need for prior knowledge of the source wavelet. A further development, the upside-down Rayleigh-Marchenko (UD-RM) method \citep{Wang2023, Wang2024}, extends the applicability of the RM method by exploiting source-receiver reciprocity. More specifically, by swapping the role of sources and receivers, the UD-RM method avoids spatial integration over receivers, making it particularly suitable for modern seabed acquisition systems where receiver sampling is usually sparse.
Compared to the original Marchenko method, both the RM and the UD-RM methods offer significant practical advantages, including the elimination of source wavelet estimation, the ability to account for surface-related multiples, and flexibility in terms of source/receiver positioning. However, these advantages come at the cost of increased data and computational demands. The RM and UD-RM methods require twice the amount of input data compared to the original Marchenko scheme, exacerbating the need for more efficient computational strategies.

In recent years, deep learning~\citep{Geoffrey2006} has emerged as an effective tool to solve inverse problems in various scientific fields. Convolutional neural networks~\citep{Lecun1998, Browne2003} have show remarkable capabilities to identifying and process complex patterns in scenarios where the input and output data are regularly sampled over 1-, 2-, or N-dimensional grids. Among various architectures that have been proposed in the literature, the U-Net architecture, originally developed for biomedical applications~\citep{Ronneberger2015}, has been a popular choice in the seismic domain~\citep{Birnie2022, Zhong2022, Sun2023, Alfarhan2024}. Building on U-Net’s demonstrated ability to process seismic data, we propose a U-Net-based approach to accelerate the UD-RM method by reducing the computational cost associated with the estimation of the focusing functions.
It is worth noting that a deep learning-based approach was proposed in \cite{Wang2025} to predict the up-going focusing functions using a U-Net architecture and subsequently compute the down-going focusing functions from their physical relationship. While effective for the original Marchenko method, this strategy is not directly applicable to the RM or UD-RM methods, where the relationship between the up- and down-going focusing functions is not direct (i.e., the down-going focusing function cannot be obtained by simply applying a forward operator to the up-going focusing function).
Therefore, our work can be seen as a follow-up to \cite{Wang2025}, in which we train a U-Net neural network on a small subset of both up- and down-going focusing functions precomputed at randomly selected imaging points. The network is tasked to learn the mapping from the initial subsurface wavefields (those commonly used for conventional imaging) to the final up- and down-going focusing functions. Once trained, the network is employed to predict focusing functions for other imaging points, significantly reducing the reliance on computationally intensive iterative solvers. 
During training, a normalized mean squared error (MSE) loss function is employed to account for amplitude differences between the up- and down-going focusing functions. Additionally, the spatial position of each imaging point is embedded into the network to enhance the learning process by informing the network of the possible changes in the wavefield characteristics according to the different positions in the domain of interest. Finally, to further improve training stability and accuracy, a time mask is applied to the predicted focusing functions, ensuring that the network focuses to predict only their physically meaningful part. These strategies promote a more balanced and accurate learning process.
Although the training process follows a supervised learning paradigm, which contains both input data and corresponding labels during training, the proposed method does not require external datasets. Instead, all training and validation samples are derived directly from the target imaging area. As a result, the proposed method retains the high accuracy typical of supervised learning while embodying the stability and adaptability of self-supervised learning (SSL) approaches, making it highly suitable for field data applications and different geological settings.

The proposed method is first validated on a synthetic model by comparing the predicted focusing functions and the resulting images with those obtained through conventional iterative methods. To further assess its robustness in sparse acquisition scenarios, a second experiment is conducted with a 40\% reduction in receiver sampling. In both cases, training is performed with progressively smaller subsets of training data and an increased number of training epochs, reducing the overall cost, which is mostly involved in preparing the training data. The results demonstrate that the proposed approach can significantly reduce the computational cost of the conventional UD-RM method while maintaining comparable imaging quality, even under sparse acquisition conditions. 
Finally, the proposed method is applied to the Volve field dataset; the resulting imaging products are shown to lead to minimal leakage compared to those obtained from the conventional UD-RM approach, therefore successfully eliminating artifacts related to multiples. These results verify the robustness and practical applicability of the proposed method to field data, highlighting its potential application for other large-scale imaging tasks.

\section*{Theory}
\subsection{A brief review of the UD-RM method}
This section provides a concise overview of the theoretical foundation of the UD-RM method; readers are referred to \cite{Wang2024} for a detailed derivation.

The UD-RM method assumes wave propagation in a heterogeneous and lossless medium~\citep{Wang2024} and involves four key depth levels, ordered from shallowest to deepest: the free surface, the source level ($\Lambda _{S}$), the receiver level ($\Lambda_{R}$), and the focal point or virtual source level ($\Lambda_{F}$). Following the wavefield notation introduced by \cite{Wang2024}, waves are always assumed to propagate toward sources and receivers: a plus sign (+) indicates a wavefield in which the waves are reaching a source or a receiver from above (referred to as down-going wavefields), while a minus sign (-) represents a wavefield in which waves are reaching from below (referred to as up-going wavefields). Using this convention, the UD-RM equations can be derived by combining the Marchenko equations with the equations derived in \cite{Almagro2014} for multi-dimensional deconvolution of data with sources placed in the subsurface (see \cite{Wang2025a} for a more detailed explanation). These equations establish the relationships between the subsurface wavefields, focusing functions, and up-/down-going separated recordings, forming the basis for the redatuming:
\begin{equation}
\label{eq:UDRM_matrix}
\begin{bmatrix}
    \mathbf{-g^{+,-}} \\
    \mathbf{g^{+,+*}}
\end{bmatrix}
=
\begin{bmatrix}
    \partial_{z}\mathbf{P}^{+,+} & \partial_{z}\widetilde{\mathbf{P}}^{+,-} \\
    \partial_{z}\widetilde{\mathbf{P}}^{+,-*} & \partial_{z}\mathbf{P}^{+,+*} 
\end{bmatrix}
\begin{bmatrix}
    \mathbf{f^{-}} \\
    \mathbf{f^{+}}
\end{bmatrix}
.
\end{equation}
where the first symbol in the superscript $^{.,.}$ identifies the wavefield component on the receiver side, while the second symbol denotes the component on the source side. The term $\mathbf{g^{+,-/+}}$ represents the up-/down-going subsurface wavefields originating from the focal point and propagating toward the receivers, with only down-going component recorded at the receiver level. Similarly, $\mathbf{f}^{-/+}$ corresponds to up-/down-going focusing functions that originate from sources to focal points. 
$\partial_{z}\mathbf{P}^{+,+}$ and $\partial_{z}\widetilde{\mathbf{P}}^{+,-}$ represent multi-dimensional convolution operators with the wavefield down-going at the receiver side and down-/up-going at the source side, while the tilde superscript indicates that the direct wave has been removed: $\partial_{z}\widetilde{\mathbf{P}}^{+,-} = \partial_{z}\mathbf{P}^{+,-} - \partial_{z}\mathbf{P}_{d}$. The operator $\partial_{z}$ denotes the vertical derivative, while the superscript $^{*}$ refers to the complex conjugate in the frequency domain or time reversal in the time domain.

Since subsurface wavefields contain only events occurring after the traveltime $t_{d}(-\textbf{x}_{R},\textbf{x}_{F})$ of the first arrival from an imaging point to mirrored receivers, their up- and down-going components can be isolated using a windowing operator $\Theta_{-x_R}$. 
This operator removes all the events occurring after $t_{d}(-\textbf{x}_{R},\textbf{x}_{F})$ as well as those before  $-t_{d}(-\textbf{x}_{R},\textbf{x}_{F})$. By applying $\Theta_{-x_R}$ on both sides of equation \ref{eq:UDRM_matrix}, the resulting system retains only two unknowns: the up-going focusing function $\mathbf{f^{-}}$ and the down-going focusing function without the direct wave $\mathbf{f^{+}_m}$. This reduced system can be solved using iterative methods such as LSQR~\citep{Ravasi2021}:
\begin{equation}
\label{eq:UDRM_matrix_window}
\begin{bmatrix}
    -\Theta_{-x_R}\partial_{z}\widetilde{\mathbf{P}}^{+,-}\mathbf{f}_\mathbf{d}^+ \\
    -\Theta_{-x_R}\partial_{z}\mathbf{P^{+,+*}}\mathbf{f}_\mathbf{d}^+
\end{bmatrix}
=
\begin{bmatrix}
    \Theta_{-x_R}\partial_{z}\mathbf{P}^{+,+} & \Theta_{-x_R}\partial_{z}\widetilde{\mathbf{P}}^{+,-} \\
    \Theta_{-x_R}\partial_{z}\widetilde{\mathbf{P}}^{+,-*} & \Theta_{-x_R}\partial_{z}\mathbf{P}^{+,+*} 
\end{bmatrix}
\begin{bmatrix}
    \mathbf{f^{-}} \\
    \mathbf{f^{+}_m}
\end{bmatrix}
,
\end{equation}
where the total down-going focusing function $\mathbf{f^{+}}$ is expressed as the sum of a direct arrival component $\mathbf{f_{d}^{+}}$, which can be computed using the migration velocity model, and its coda $\mathbf{f_{m}^{+}}$.

In the presence of sparse receiver geometries, estimating the focusing functions becomes an under-determined inverse problem. To mitigate illumination artifacts associated with this under-determination, a sparsity-promoting inversion with a sliding linear Radon sparsifying transform $\mathbf{S}$ can be applied to recover the focusing functions~\citep{Wang2024}:
\begin{equation}
\label{eq:sparse}
\min_{\mathbf{z}_f}|| \mathbf{d}-\mathbf{G}\mathbf{S}^{H}\mathbf{z}_f||_{2}^{2}+\lambda || \mathbf{z}_f|| _{1}
,
\end{equation}
where the vector $\mathbf{z}_f=[\mathbf{z}_{f^-}^T, \mathbf{z}_{f_m^+}^T]^T$ contains the sparse representation of the up-going focusing function and the coda of the down-going focusing functions. The terms $\mathbf{d}$ and $\mathbf{G}$ identify the left-hand side vector and the operator in equation \ref{eq:UDRM_matrix_window}, respectively. The sparsifying transform $\mathbf{S}$ consists of a sliding linear Radon transform \citep{Rickett2014}, which performs a linear Radon transform within a moving window to extract local planar features from the seismic wavefield. The resulting sparse inversion problem can be efficiently solved using the Fast Iterative Shrinkage-Thresholding Algorithm (FISTA) \citep{Beck2009}. Once the focusing functions are obtained, the corresponding subsurface wavefields can be reconstructed by applying equation \ref{eq:UDRM_matrix}.

\begin{figure*}[!t]
\centering
\includegraphics[width=1\textwidth]{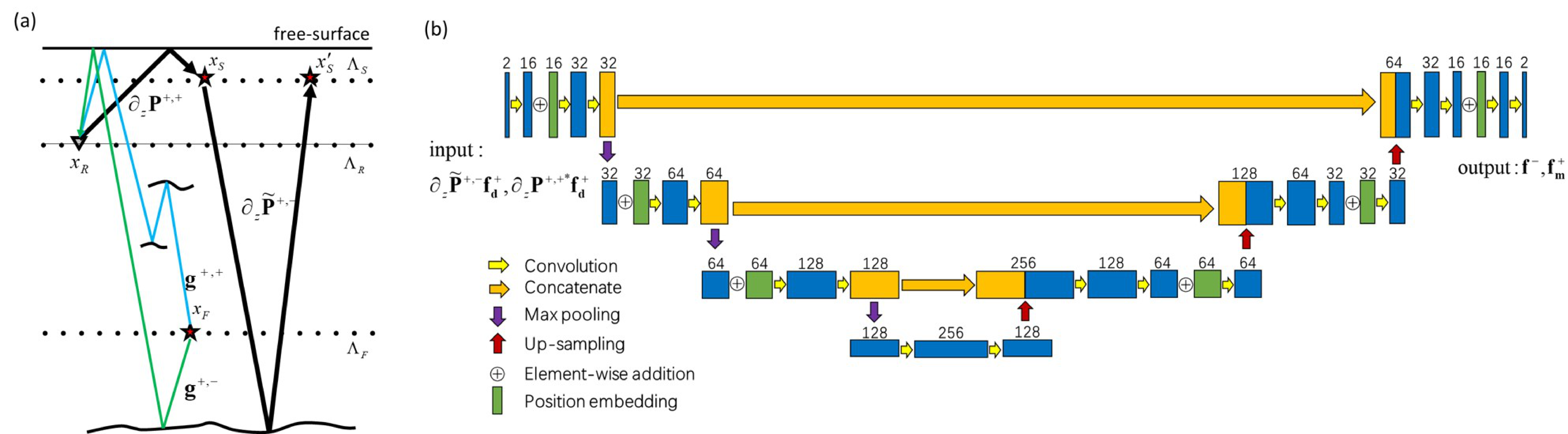}
\caption{(a) Schematic representation of the wavefields involved in the UD-RM equations, where the black ray represents the recorded wavefield, the green ray represents the up-going component of the subsurface wavefield, and the blue ray represents the down-going component of the subsurface wavefield. (b) The architecture of U-Net used for focusing function prediction.}
\label{fig:geometries}
\end{figure*} 

\subsection{Deep Learning for focusing function estimation}
In this work, we propose a deep learning-based solution as a way to accelerate the computation of focusing functions. As demonstrated in \cite{Wang2024}, the initial subsurface wavefields, which neglect the contributions of surface-related and internal multiples, can be obtained using the initial estimate of the focusing functions ($\mathbf{f}^{+}=\mathbf{f}_\mathbf{d}^{+}$, $\mathbf{f}^{-}=0$). This leads to the initial up- and down-going subsurface wavefields expressed as $\mathbf{-g^{+,-}_0}=\partial_{z}\widetilde{\mathbf{P}}^{+,-}\mathbf{f}_\mathbf{d}^+$ and $\mathbf{g^{+,+*}_0}=\partial_{z}\mathbf{P^{+,+*}}\mathbf{f}_\mathbf{d}^+$, respectively. Leveraging these wavefields, we propose to train a neural network to map the initial subsurface wavefields ($\mathbf{-g^{+,-}_0}$ and $\mathbf{g^{+,+*}_0}$) to the final focusing functions ($\mathbf{f^{-}}$ and $\mathbf{f^{+}_m}$). The network is trained on a small subset of focusing functions, precomputed at randomly selected imaging points within the target area, serving as a proxy for solving equation~\ref{eq:UDRM_matrix_window}, or equation~\ref{eq:sparse} in cases of sparse receiver arrays. After training, the network is applied to predict focusing functions at all other imaging locations.
This framework focuses on predicting the focusing functions rather than the subsurface wavefields directly, as the focusing functions are confined within a shorter time window, making them inherently easier to model and learn. The predicted focusing functions are then used to reconstruct the subsurface wavefields through equation~\ref{eq:UDRM_matrix}, which are ultimately employed for imaging.

The neural network employed in this study is a standard U-Net architecture~\citep{Ronneberger2015}, composed of a series of contracting blocks in the encoder and corresponding expanding blocks in the decoder, connected through skip connections (Figure \ref{fig:geometries}b). To enhance the network stability and generalization, we use Leaky ReLU activation functions with a negative slope of 0.2, batch normalization, and $50\%$ dropout are applied after each convolutional layer.
To enable the network to better capture the spatial variations in wavefield characteristics across different imaging points, spatial position information of each imaging point is embedded into both the encoder and decoder paths of the U-Net. Specifically, the lateral and vertical coordinates of the imaging point are first normalized to the range [–1, 1]. They are then passed through a sinusoidal positional embedding module, where sine and cosine functions at multiple frequencies are used to generate high-dimensional embeddings. These embeddings are then processed by small multilayer perceptron and concatenated to form the full positional embedding vector.
This vector is injected into every level of the U-Net via learned linear projections that adapt it to the appropriate number of feature channels at that level. In each encoder and decoder block, the embedding is broadcast to match the spatial dimensions of the feature maps and added directly to the convolutional features. This strategy allows the network to learn location-aware transformations without increasing the number of input channels, improving the network’s ability to generalize focusing function predictions to unseen imaging points.
The network is optimized using an Adam optimizer~\citep{Kingma2014}, with the loss function defined as the normalized mean squared error (MSE) between the predicted and pre-computed focusing functions:
\begin{equation}
\label{eq:loss}
\mathcal{L} = \frac{|| \Theta_{-x_R}(\mathbf{\hat{f}}^{-} - \mathbf{f}^{-}) ||_2^2}{|| \mathbf{f}^{-} ||_2^2} + \frac{|| \Theta_{-x_R}(\mathbf{\hat{f}}^{+} - \mathbf{f}^{+}) ||_2^2}{|| \mathbf{f}^{+} ||_2^2},
\end{equation}
where $\mathbf{\hat{f}}^{-/+}$ is the predicted up- and down-going focusing functions, while $\mathbf{f}^{-/+}$ represents their pre-computed target counterparts. $||\cdot||_2^2$ denotes the squared L2 norm. Normalizing the loss by the energy of the true focusing functions prevents the loss focusing only on higher-amplitude components, ensuring balanced gradient updates during training.
Furthermore, a time window operator $\Theta_{-x_R}$ is applied prior to the loss calculation to restrict the network’s attention to the physically meaningful parts of the focusing functions, which occur before the first arrival traveltime from the focal point to the mirrored receivers or after the corresponding negative traveltime. Incorporating such a physical constraint into the learning process leads to more physically consistent and reliable predictions.

In summary, for each imaging task, a subset of points within the target imaging area is randomly selected for training and validation, with their focusing functions pre-computed using the conventional iterative scheme. The trained network is then applied to predict the final focusing functions for the remaining points.
These predicted focusing functions are subsequently used to reconstruct the up- and down-going subsurface wavefields via equation~\ref{eq:UDRM_matrix}, which are then employed for seismic imaging.

\section*{Synthetic Examples}
\subsection{Dense receiver array}
\begin{figure*}[!t]
\centering
\includegraphics[width=1\textwidth]{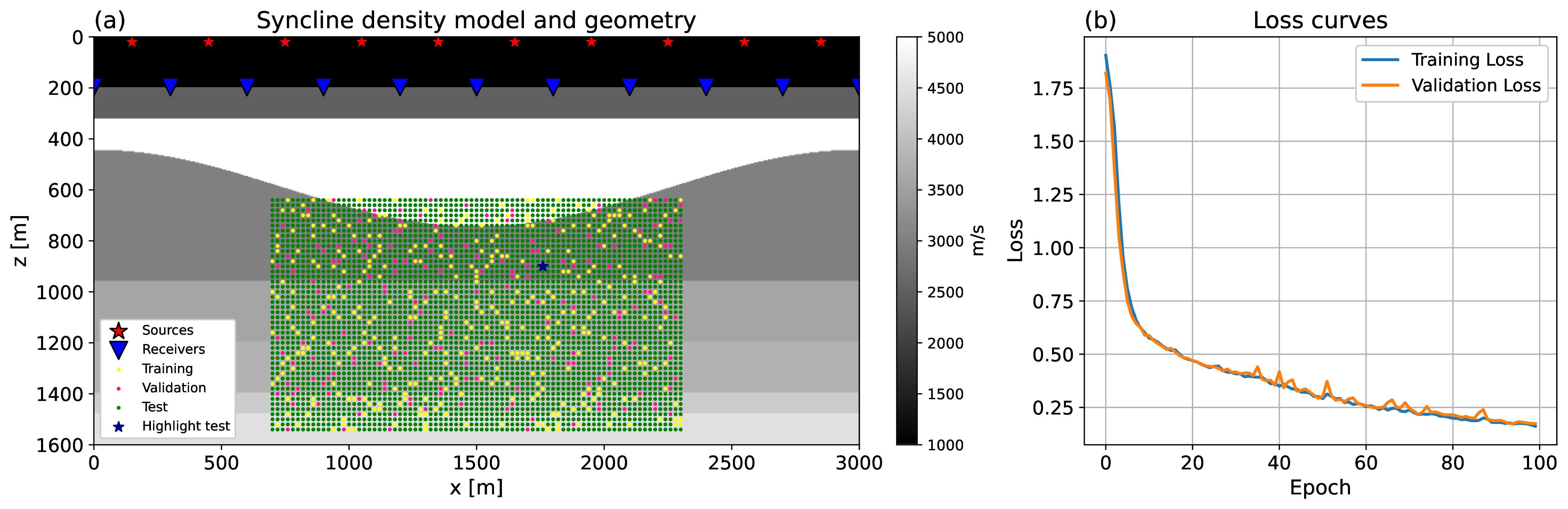}
\caption{(a) Syncline density model and geometry, (b) training and validation loss curves.}
\label{fig:geometry_simple}
\end{figure*}

To assess the effectiveness of the proposed approach, we first conduct a test using a synthetic dataset generated in a constant-velocity (2400 m/s), variable-density model (Figure \ref{fig:geometry_simple}a). The dataset is simulated with free-surface effects and includes 201 sources positioned at a depth of 20 m (red stars in Figure \ref{fig:geometry_simple}a) and 201 receivers placed at a depth of 196 m (blue triangles in Figure \ref{fig:geometry_simple}a), both uniformly distributed along the horizontal axis from 0 to 3000 m. The chosen imaging region contains 3726 imaging points, which are distributed on a uniform 20 m by 20 m grid, as shown in Figure~\ref{fig:geometry_simple}a.

The employed U-Net architecture consists of five levels of down-sampling and up-sampling blocks, with channel sizes progressively increasing from 16 to 256 (i.e., by a power of two after each dowsampling step). The network is trained with a batch size of 16 and a learning rate of 0.001. For training, 10\% of the imaging points are randomly selected as training data (yellow points), and 5\% are used as validation data (pink points), with their corresponding focusing functions computed using the LSQR solver. The remaining 85\% (green points) serve as test data, where the focusing functions are predicted using the trained network. Training is carried out on a single A100 GPU for 100 epochs, with the convergence curves of training and validation losses shown in Figure \ref{fig:geometry_simple}b.

\begin{figure*}[!t]
\centering
\includegraphics[width=1\textwidth]{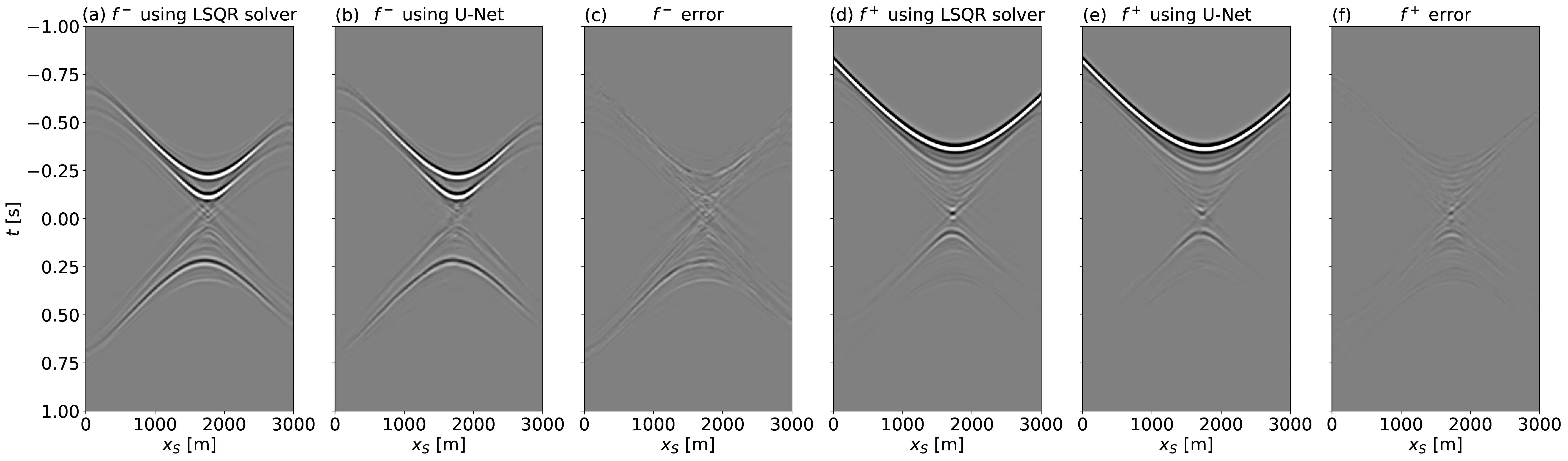}
\caption{Up-going focusing functions obtained using (a) the LSQR solver, and (b) U-Net, with their difference in (c). Down-going focusing functions obtained using (d) the LSQR solver, and (e) U-Net, with their difference in (f).}
\label{fig:f1_inv}
\end{figure*}

\begin{figure*}[!t]
\centering
\includegraphics[width=1\textwidth]{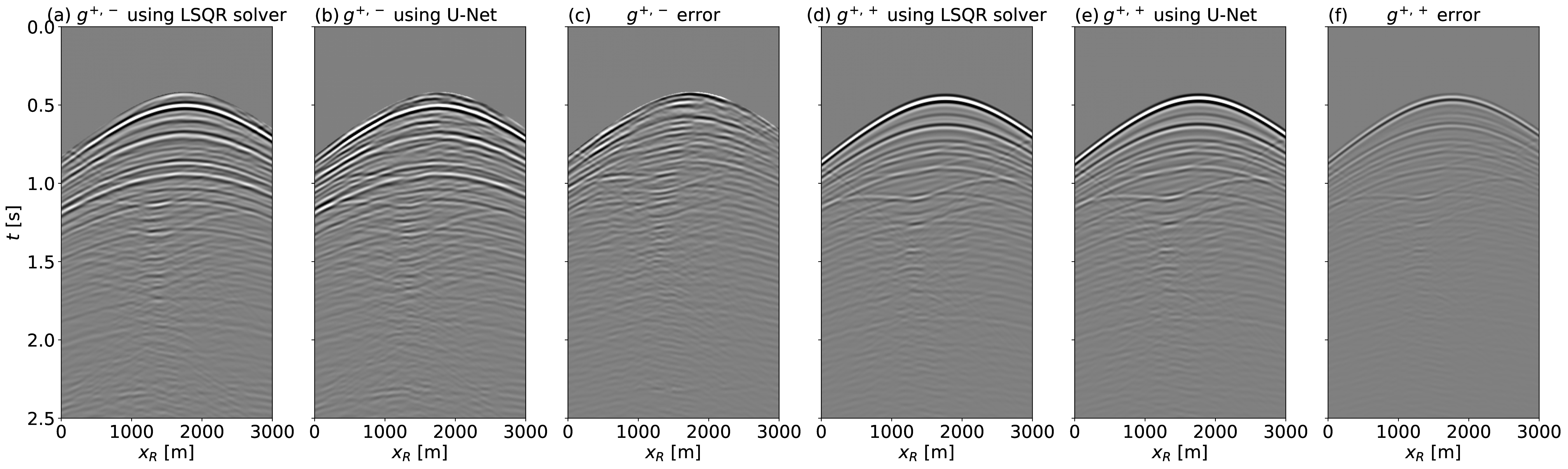}
\caption{Up-going subsurface wavefields obtained using (a) the LSQR solver, and (b) U-Net, with their difference in (c). Down-going subsurface wavefields obtained using (d) the LSQR solver, and (e) U-Net, with their difference in (f).}
\label{fig:GF}
\end{figure*}

To evaluate the performance of the trained network, we show the predicted up- and down-going focusing functions at a representative imaging point, indicated by the blue star in Figure \ref{fig:geometry_simple}a. The predicted results are shown in Figures \ref{fig:f1_inv}b and \ref{fig:f1_inv}e, while the corresponding reference solutions, computed using the LSQR solver, are displayed in Figures \ref{fig:f1_inv}a and \ref{fig:f1_inv}d. As highlighted by the difference plots in Figures \ref{fig:f1_inv}c and \ref{fig:f1_inv}f, the predicted focusing functions show excellent agreement with the reference solutions, with only minor signal leakage. To further validate the predictions, we use both the predicted and reference focusing functions to reconstruct the up- and down-going subsurface wavefields via equation \ref{eq:UDRM_matrix}. The resulting wavefields and their differences are shown in Figure \ref{fig:GF}. Due to the significantly higher energy (approximately 80 times) of the down-going wavefields compared to the up-going ones, different amplitude clipping values are applied for visualization. Consequently, although the relative error in the up-going wavefields (Figure \ref{fig:GF}c) may appear pronounced, it remains small in absolute terms.

\begin{figure*}[!t]
\centering
\includegraphics[width=1\textwidth]{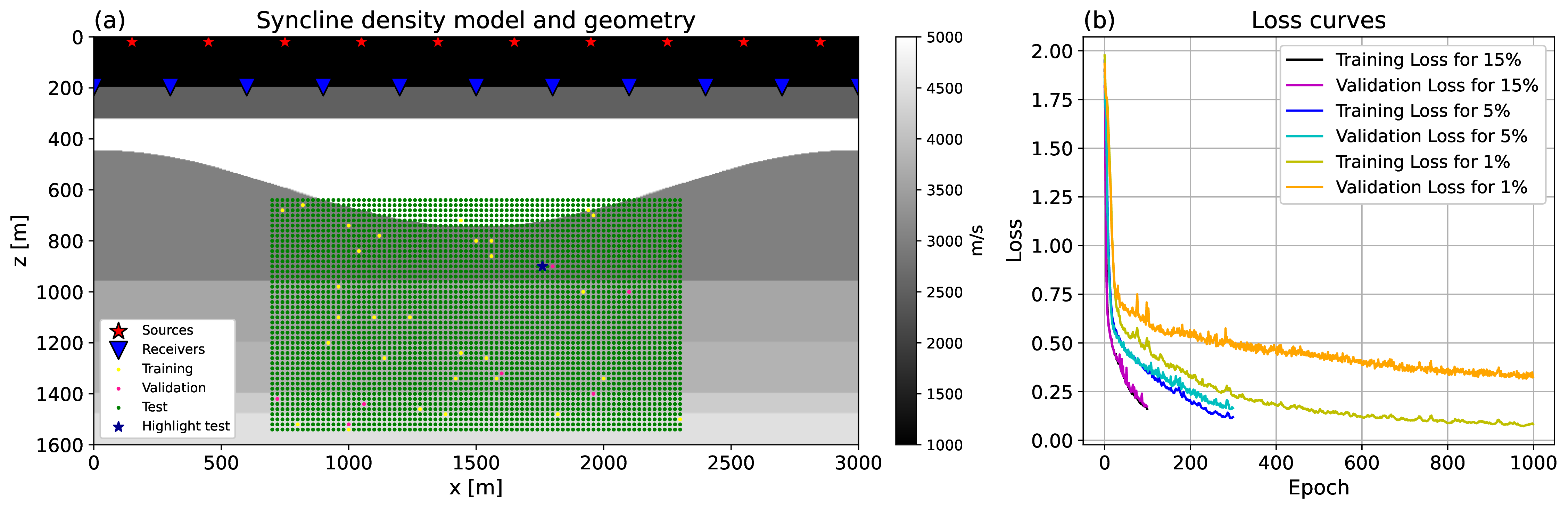}
\caption{(a) Syncline density model and acquisition geometry, showing the distribution of training (0.8\%), validation (0.2\%), and test (99\%) imaging points. (b) Training and validation loss curves for all three tests with varying amounts of training data.}
\label{fig:geometry_simple_1percent}
\end{figure*}

\begin{figure*}[!t]
\centering
\includegraphics[width=1\textwidth]{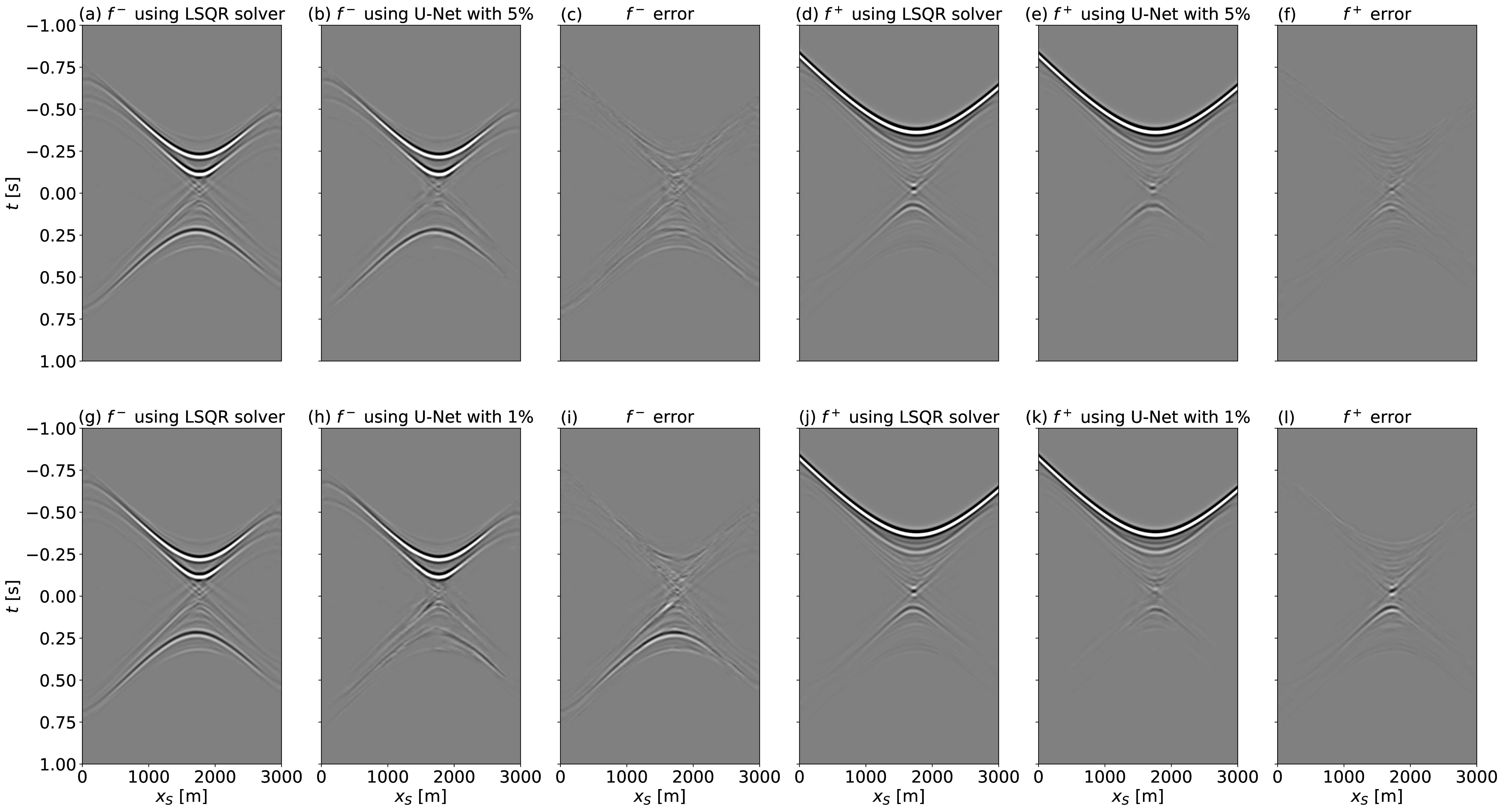}
\caption{Up-going focusing functions computed by the LSQR solver (a, g) and U-Net (b, h) with 5\% and 1\% training data; corresponding differences are shown in (c) and (i). Down-going focusing functions computed by the LSQR solver (d, j) and U-Net (e, k) with 5\% and 1\% training data; corresponding differences are shown in (f) and (l).}
\label{fig:f1_inv_1percent}
\end{figure*}

To further evaluate the trade-off between computational efficiency and prediction accuracy, we conduct two additional tests using a reduced number of training and validation imaging points: 4\% and 1\%, as well as 0.8\% and 0.2\%, respectively. As the size of the training dataset decreases, the number of training epochs is increased to 300 and 1000 to ensure sufficient gradient updates for convergence. To illustrate the difficulty of the task, Figure~\ref{fig:geometry_simple_1percent}a displays the distribution of imaging points for the most extreme case, where only 0.8\% of the points (yellow points) are used for training and 0.2\% (pink points) for validation. Figure~\ref{fig:geometry_simple_1percent}b shows the training and validation loss curves for all three cases. As expected, the overall performance degrades as fewer training samples are used, despite more training epochs are employed. In particular, for the case with 0.8\% of training data, the network exhibits a reduced generalization capability, as indicated by the gap between training and validation losses. This performance drop is attributed to insufficient spatial coverage and limited feature diversity in the training set, which restrict the network’s ability to learn representative patterns. The impact of limited training data is further illustrated by comparing the predicted and LSQR-computed focusing functions at the same imaging point marked by a blue star in Figure~\ref{fig:geometry_simple_1percent}a. These results, along with their differences, are presented in Figure~\ref{fig:f1_inv_1percent}. 
The results demonstrate that increasing the number of training epochs can partially improve the network's performance when the percentage of training imaging points is reduced. However, this increase in training duration cannot fully compensate for the accuracy loss caused by extremely limited training data, such as the 0.8\% case illustrated here. Notable prediction errors, particularly in the form of signal leakage, are observed in the difference plots shown in Figures~\ref{fig:f1_inv_1percent}i and~\ref{fig:f1_inv_1percent}l.

\begin{figure*}[!t]
\centering
\includegraphics[width=1\textwidth]{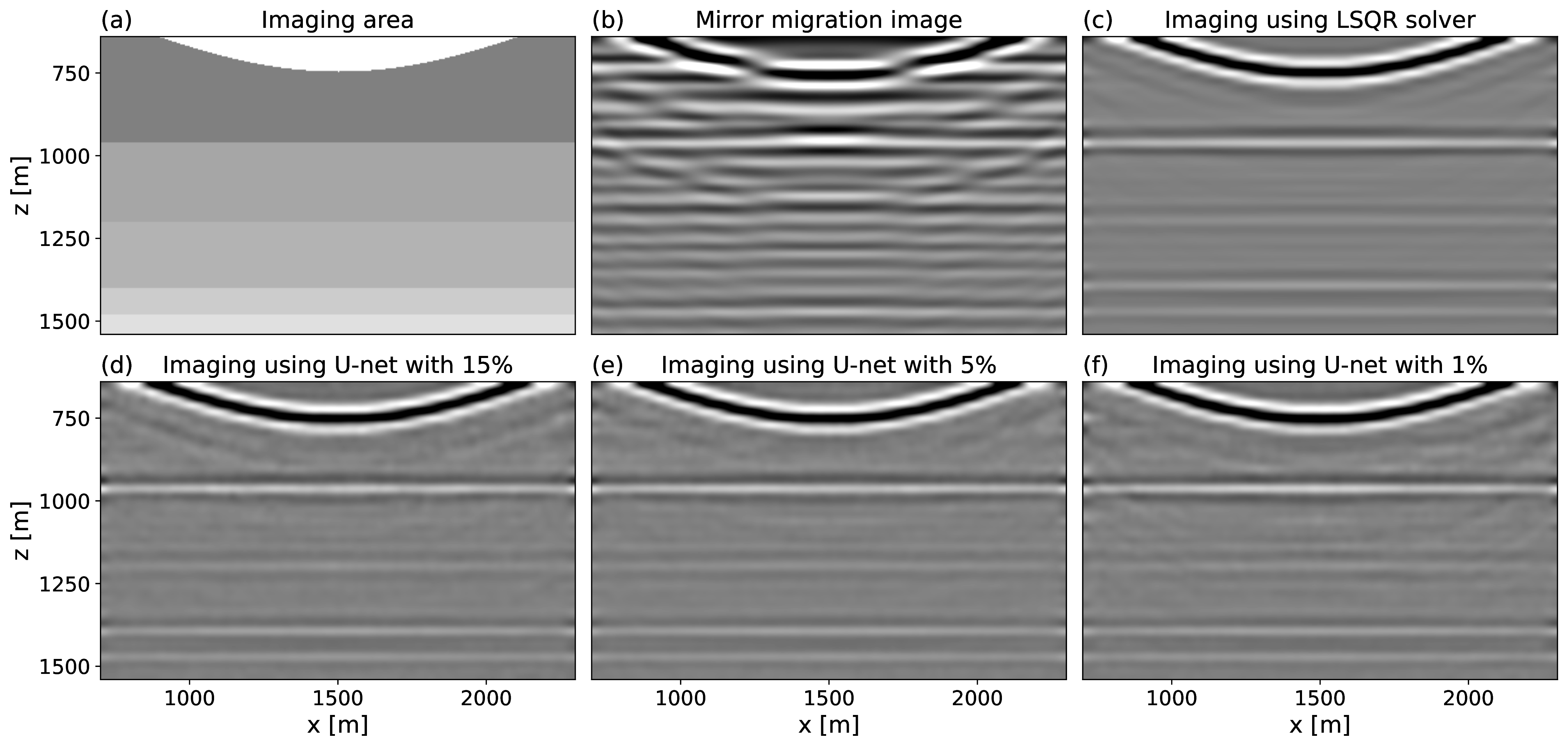}
\caption{(a) Target imaging area; (b) mirror migration result; (c) UD-RM image using the LSQR solver; (d)–(f) UD-RM images using U-Net with 15\%, 5\%, and 1\% training and validation data, respectively.}
\label{fig:Images_dense}
\end{figure*}

Subsequently, the imaging is carried out within the target area shown in Figure \ref{fig:Images_dense}a. For comparison, mirror migration, which utilizes waves propagating from mirrored receivers to imaging points under the single-scattering assumption, is also used, with results displayed in Figure \ref{fig:Images_dense}b. Due to the strong surface-related multiples, false reflectors appear in the mirror migration result. In contrast, the UD-RM image constructed using subsurface wavefields computed by the LSQR solver (Figure \ref{fig:Images_dense}c) accurately reconstructs the underlying density model, free from multiple-related artifacts. Figures \ref{fig:Images_dense}d to \ref{fig:Images_dense}f present the images obtained using the U-Net trained with 15\%, 5\%, and 1\% of the imaging points, respectively, which closely resemble the LSQR-based UD-RM image. Although some high-frequency incoherent noise is visible in the U-Net-based images, surface-related and internal multiples are successfully suppressed. 
These results demonstrate that the proposed method can reliably recover subsurface structures and correctly handle surface-related and internal multiples, achieving imaging quality comparable to that of the traditional UD-RM method, even with limited training data.

Finally, we compare the computational cost of the traditional UD-RM method using the LSQR solver with that of the proposed U-Net-based prediction approach. The LSQR solver requires approximately 1 minute 12 seconds per imaging point, resulting in a total computation time of around 74 hours 31 minutes 12 seconds for the entire imaging area. In contrast, the U-Net-based method reduces the computational time to a small fraction of that required by the conventional UD-RM approach. 
For the case using 15\% of imaging points for training and validation, the total runtime is reduced to approximately 11 hours 57 minutes 55 seconds, which includes 11 hours 10 minutes 41 seconds to calculate the focusing functions for 15\% of imaging points, 46 minutes 9 seconds for network training, and 1 minute 5 seconds for predicting the focusing functions at the remaining locations; this amounts to an approximately 6-fold reduction in computational time when taking into account the entire process.
When using 5\% of imaging points for training and validation, the total runtime further decreases to approximately 4 hours 28 minutes 58 seconds. This comprises 3 hours 43 minutes 34 seconds for computing focusing functions, 44 minutes 12 seconds for training, and 1 minute 12 seconds for predicting, yielding an 18-fold speed-up.
With only 1\% of the imaging points used for training and validation, the total runtime drops to approximately 1 hour 20 minutes 5 seconds, including 44 minutes 43 seconds for focusing function computation, 34 minutes 10 seconds for training, and 1 minutes 12 seconds for predicting, resulting in approximately a 56-fold improvement in computational efficiency.
These results demonstrate that the proposed U-Net-based method greatly reduces the computational cost of the UD-RM method while maintaining image quality comparable to the conventional one. To prioritize computational efficiency while maintaining acceptable accuracy, we can reduce the number of training points while compensating with additional training epochs.

\subsection{Sparse receiver array}
\begin{figure*}[!t]
\centering
\includegraphics[width=1\textwidth]{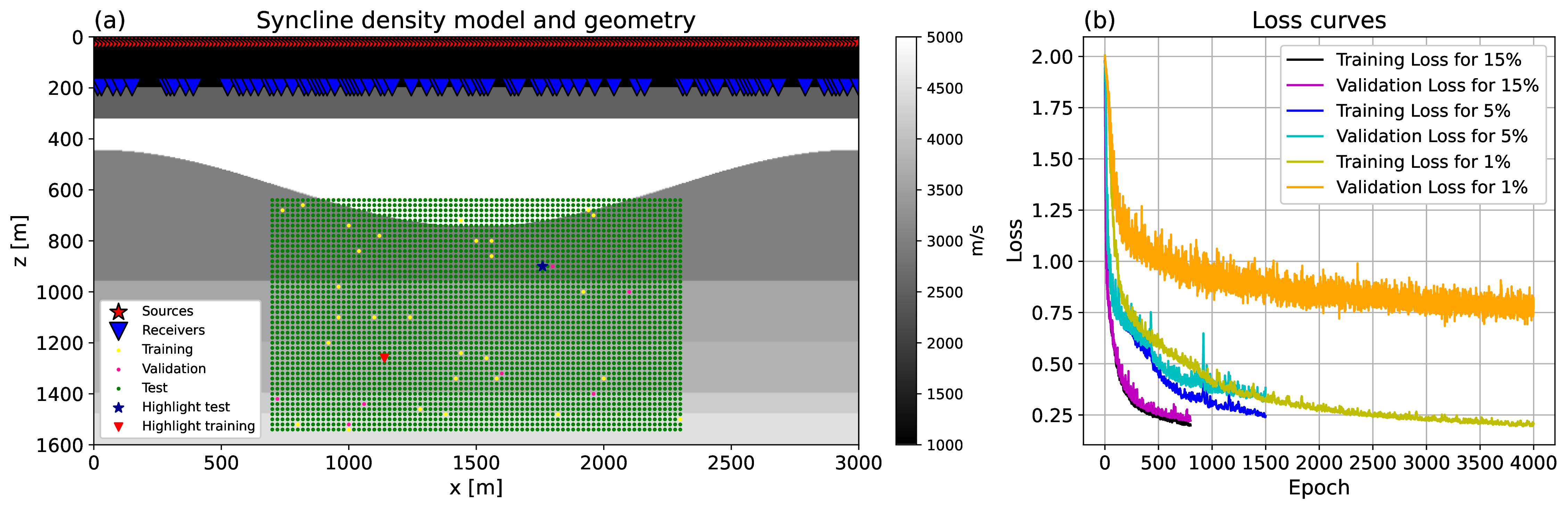}
\caption{(a) Syncline model with sparse receiver array, showing the distribution of training (0.8\%), validation (0.2\%), and test (99\%) imaging points. (b) Training and validation loss curves for tests with sparse receiver array.}
\label{fig:geometry_simple_sparse}
\end{figure*}

To assess the robustness of the proposed method under sparse acquisition conditions, we randomly retain 40\% of the total receivers, as shown in Figure~\ref{fig:geometry_simple_sparse}a. As discussed in the Theory section, reducing receiver coverage transforms equation~\ref{eq:UDRM_matrix_window} into a heavily under-determined inverse problem. To enhance the quality of the focusing functions under these conditions, the FISTA solver needs 200 iterations to solve equation \ref{eq:sparse}. Similarly, to improve the computational efficiency, the U-Net is trained to approximate the sparse inversion process in equation \ref{eq:sparse}, mapping from the up- and down-going subsurface wavefields to the pre-computed up- and down-going focusing functions.

\begin{figure*}[!t]
\centering
\includegraphics[width=0.8\textwidth]{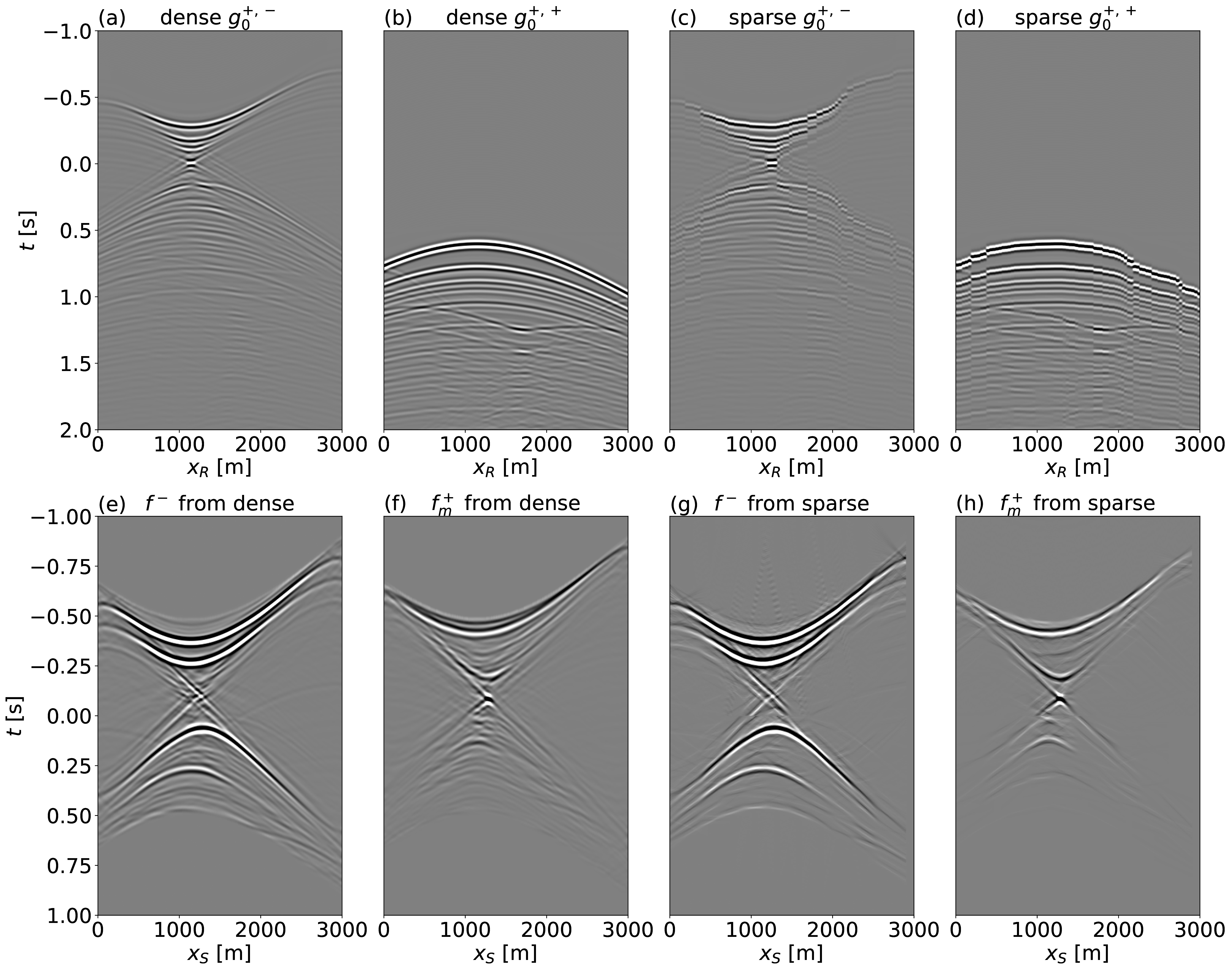}
\caption{(a, b) Initial up- and down-going subsurface wavefields for the dense receiver array; (c, d) initial up- and down-going subsurface wavefields for the sparse receiver array; (e, f) up- and down-going focusing functions retrieved using the dense receiver array; (g, h) up- and down-going focusing functions retrieved using the sparse receiver array.}
\label{fig:in_out_put_data}
\end{figure*}

As illustrated in the Theory section, the subsurface waves propagate from the focal point to the mirrored receivers, while focusing functions originate from the sources to the focal point. 
Accordingly, for the sparse receiver test, the input of the network, shown in Figures \ref{fig:in_out_put_data}c and \ref{fig:in_out_put_data}d and corresponding to the training imaging points marked by red triangles in Figure~\ref{fig:geometry_simple_sparse}a, is a sparsely sampled version of the input used in the dense receiver case (Figures~\ref{fig:in_out_put_data}a and \ref{fig:in_out_put_data}b).
The target outputs in the dense receiver scenario are computed using densely sampled data and solved via equation~\ref{eq:UDRM_matrix_window} (Figures~\ref{fig:in_out_put_data}e and \ref{fig:in_out_put_data}f), while the outputs for the sparse case are generated using sparsely sampled data and obtained by solving equation~\ref{eq:sparse} (Figures~\ref{fig:in_out_put_data}g and \ref{fig:in_out_put_data}h). The latter exhibits slightly reduced accuracy due to the limited receiver coverage.
Because the input under sparse sampling contains only 40\% of the original spatial information while the output size remains unchanged, an up-sampling layer is added to the end of the U-Net to interpolate the feature maps and match the output dimensions. Apart from this modification, all other network parameters remain consistent with those used in the dense receiver experiments.

\begin{figure*}[!t]
\centering
\includegraphics[width=1\textwidth]{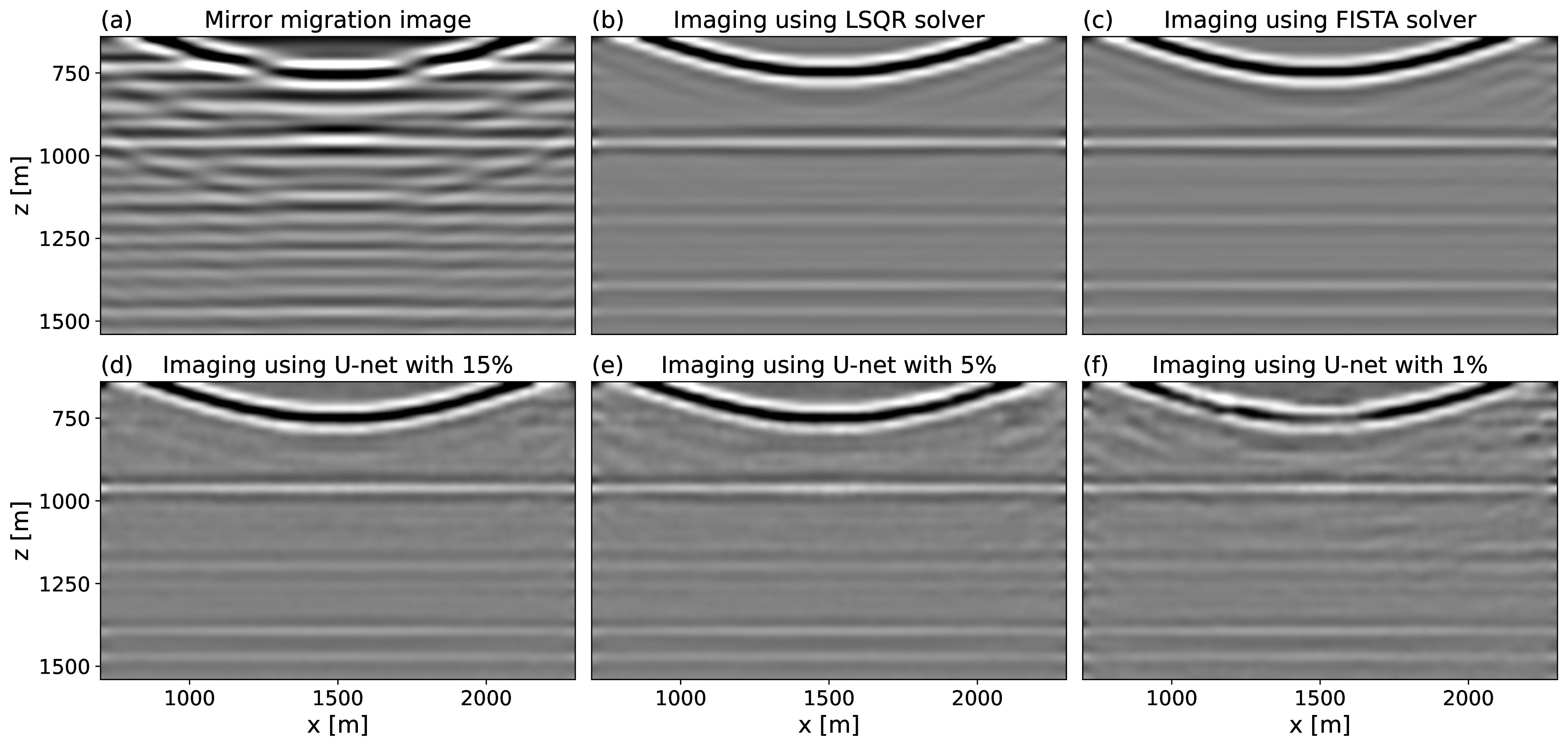}
\caption{(a) Mirror migration result; (b) UD-RM image using LSQR with full receiver array;  
(c) UD-RM image using FISTA with 40\% receiver coverage; (d–f) Sparse UD-RM images using U-Net trained with 15\%, 5\%, and 1\% of training and validation data, respectively.}
\label{fig:Images_sparse}
\end{figure*}

\begin{figure*}[!t]
\centering
\includegraphics[width=0.75\textwidth]{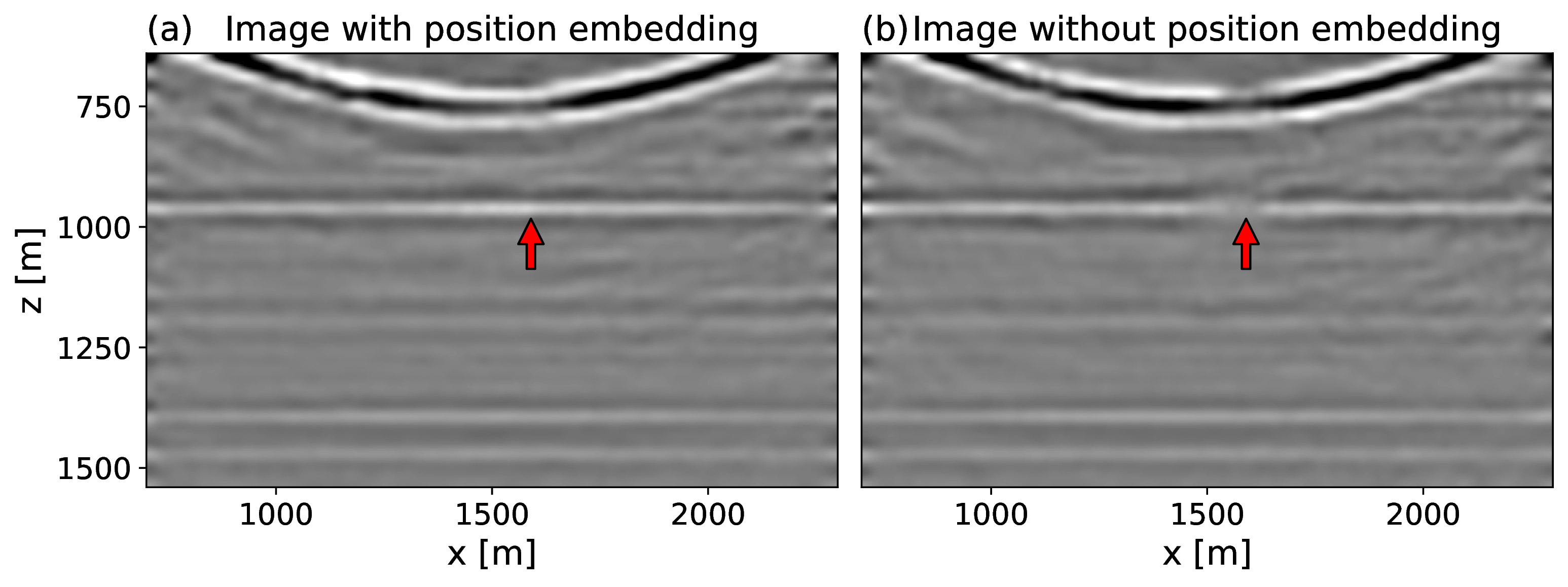}
\caption{Sparse UD-RM images using U-Net trained with 1\% of training and validation data (a) with image position embedding and (b) without image position embedding.}
\label{fig:Images_sparse_without_position}
\end{figure*}

We evaluate the proposed approach under sparse receiver conditions by conducting three experiments using 10\% and 5\%, 4\% and 1\%, and 0.8\% and 0.2\% of the imaging points for training and validation, respectively. 
Due to the reduced input size (40\% of the dense case), we increase the number of training epochs to 800, 1500, and 4000 to ensure stable convergence in this more challenging setting. The training and validation loss curves for these tests are presented in Figure~\ref{fig:geometry_simple_sparse}b.
Following training, the networks are used to predict the remaining focusing functions, which are then used to reconstruct the subsurface wavefields via equation~\ref{eq:UDRM_matrix}. The resulting image computed from wavefields obtained using the FISTA solver (Figure~\ref{fig:Images_sparse}c) closely resembles the image derived from the full receiver array using the LSQR solver (Figure~\ref{fig:Images_sparse}b), demonstrating the effectiveness of sparsity-promoting inversion to handle sparse acquisition.
Images generated using U-Net-predicted subsurface wavefields for each training data percentage are displayed in Figures \ref{fig:Images_sparse}d to \ref{fig:Images_sparse}f. As the number of training samples decreases, the image quality degrades, with increased high-frequency noise visible, particularly in Figure~\ref{fig:Images_sparse}f. 
Nevertheless, all U-Net-based results successfully capture the subsurface structure and effectively suppress artifacts associated with surface-related and internal multiples, which are typically difficult to handle by the mirror migration (as shown in Figure~\ref{fig:Images_sparse}a).
To assess the impact of incorporating imaging point coordinates into the network, we compare two sparse UD-RM image using the U-Net trained with 1\% of data. As shown in Figure~\ref{fig:Images_sparse_without_position}b, the model trained without positional information exhibits noticeable discontinuities, particularly in the region highlighted by the red arrow. In contrast, Figure~\ref{fig:Images_sparse_without_position}a shows that incorporating positional embedding leads to a more continuous and geologically realistic structure in the same area.
These results confirm that the proposed approach is capable of producing reliable images under sparse acquisition conditions, maintaining robustness and imaging quality even with minimal training data.

It is important to note that, sparse inversion algorithms typically exhibit slower convergence than their least-squares counterparts, resulting in significantly higher computational costs. For this test with a sparse receiver geometry, the FISTA solver requires 200 iterations to converge to a satisfactory solution, leading to approximately 10 minutes 47 seconds per imaging point. Therefore, the total computation time is around 669 hours 38 minutes 42 seconds for the entire imaging area. 
In this context, the U-Net-based approach becomes even more important to reduce the computational burden to a practical level. 
When using 15\% of the imaging points for training and validation, the total runtime is reduced to approximately 103 hours 24 minutes 57 seconds. This includes 100 hours 26 minutes 48 seconds for computing focusing functions, 2 hours 58 minutes for training, and 32 seconds for inference to predict focusing functions at the remaining points, yielding a 6-fold reduction in total computation time.
With 5\% training and validation data, the total runtime decreases to approximately 35 hours 19 minutes 36 seconds. This comprises 33 hours 28 minutes 56 seconds for focusing function computation, 1 hour 50 minutes 8 seconds for training, and 31 seconds for inference, resulting in a 19-fold speed-up.
The most efficient case uses only 1\% of imaging points for training and validation, reducing the total computation time to approximately 7 hours 39 minutes 18 seconds, which includes 6 hours 41 minutes 47 seconds for focusing function computation, 56 minutes 59 seconds for training, and 32 seconds for inference. This achieves an 87-fold reduction in computational time.
These results demonstrate that the proposed U-Net-based method can address the computational challenges of solving the UD-RM equations by sparse inversion. Even though there is an inherent trade-off between imaging quality and computational cost, all results show that the proposed method effectively handles sparse receiver geometries and suppresses both surface-related and internal multiples.

\section*{Field Example}
\begin{figure*}[!t]
\centering
\includegraphics[width=1\textwidth]{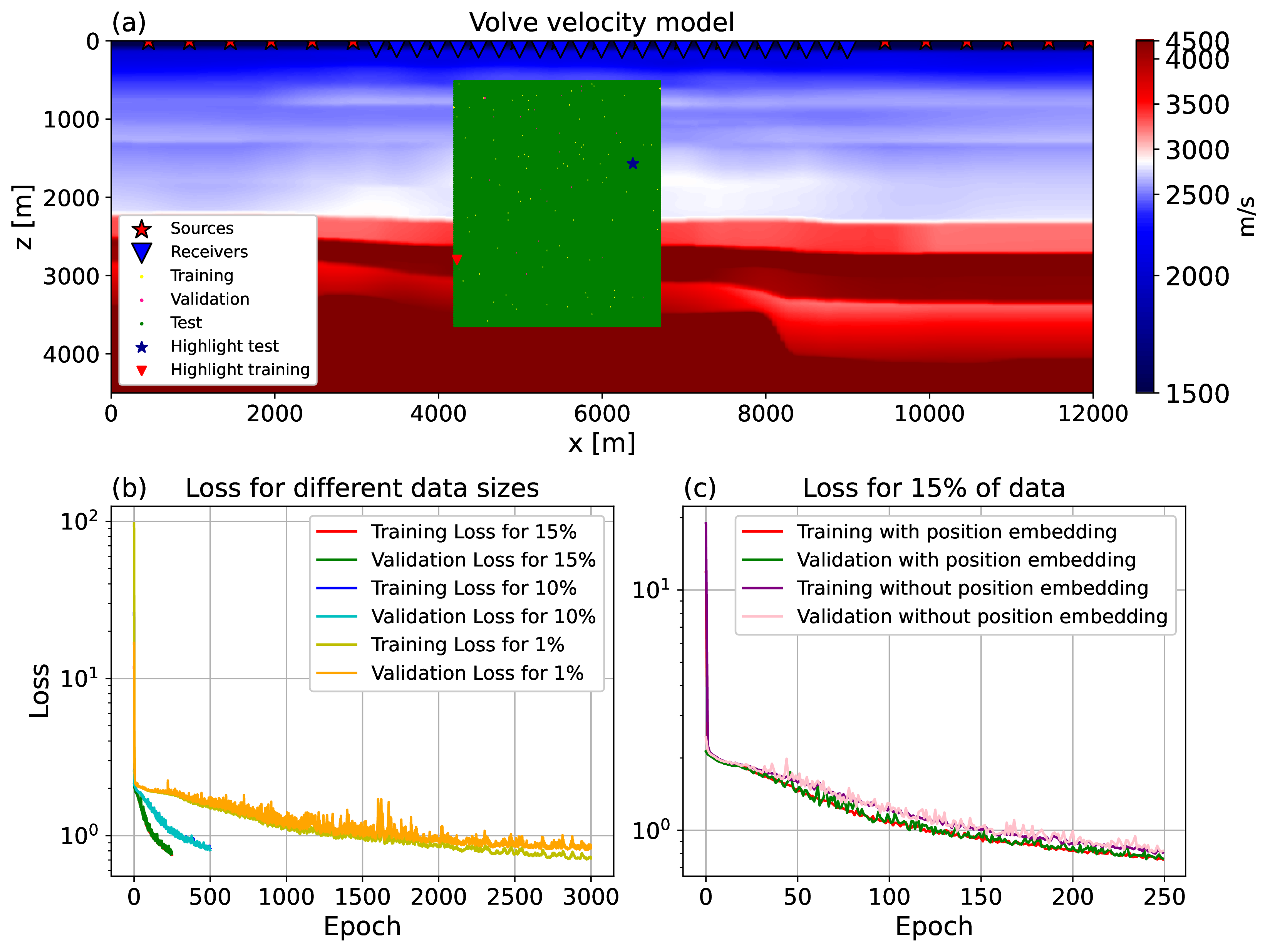}
\caption{(a) 2D migration velocity model and acquisition geometry, showing the distribution of training (0.8\%), validation (0.2\%), and test (99\%) imaging points. (b) Training and validation loss curves for different data sizes. (c) Training and validation loss curves for 15\% of the data with and without position embedding.}
\label{fig:geometry_volve}
\end{figure*}

We now apply the proposed method to the Volve field dataset, which was acquired over the Volve field, located in the central North Sea, offshore Norway~\citep{Szydlik2007}. For this study, we select a 2D receiver line consisting of 240 receivers and an adjacent line of 240 sources. The receivers are positioned at a depth of approximately 92 m with a spacing of 25 m, ranging from 3012 to 8996 m. The sources are located approximately 9.6 m below the free surface, spaced every 50 m, covering a range from 6 to 11955 m. 
A 2D migration velocity slice, shown in Figure \ref{fig:geometry_volve}a, is extracted from the 3D migration velocity model along the line closest to the selected receiver and source locations. A total of 10605 imaging points are chosen within this slice, arranged on a grid with 30 m vertical and 25 m horizontal spacing, covering depths from 520 to 3640 m. 
Following the same procedure used in the synthetic examples, we conduct three experiments using different proportions of imaging points for training and validation: 15\% (10\% training, 5\% validation), 5\% (4\% training, 1\% validation), and 1\% (0.8\% training, 0.2\% validation). The spatial distribution of imaging points for the 1\% case is shown in Figure~\ref{fig:geometry_volve}a.

\begin{figure*}[!t]
\centering
\includegraphics[width=0.8\textwidth]{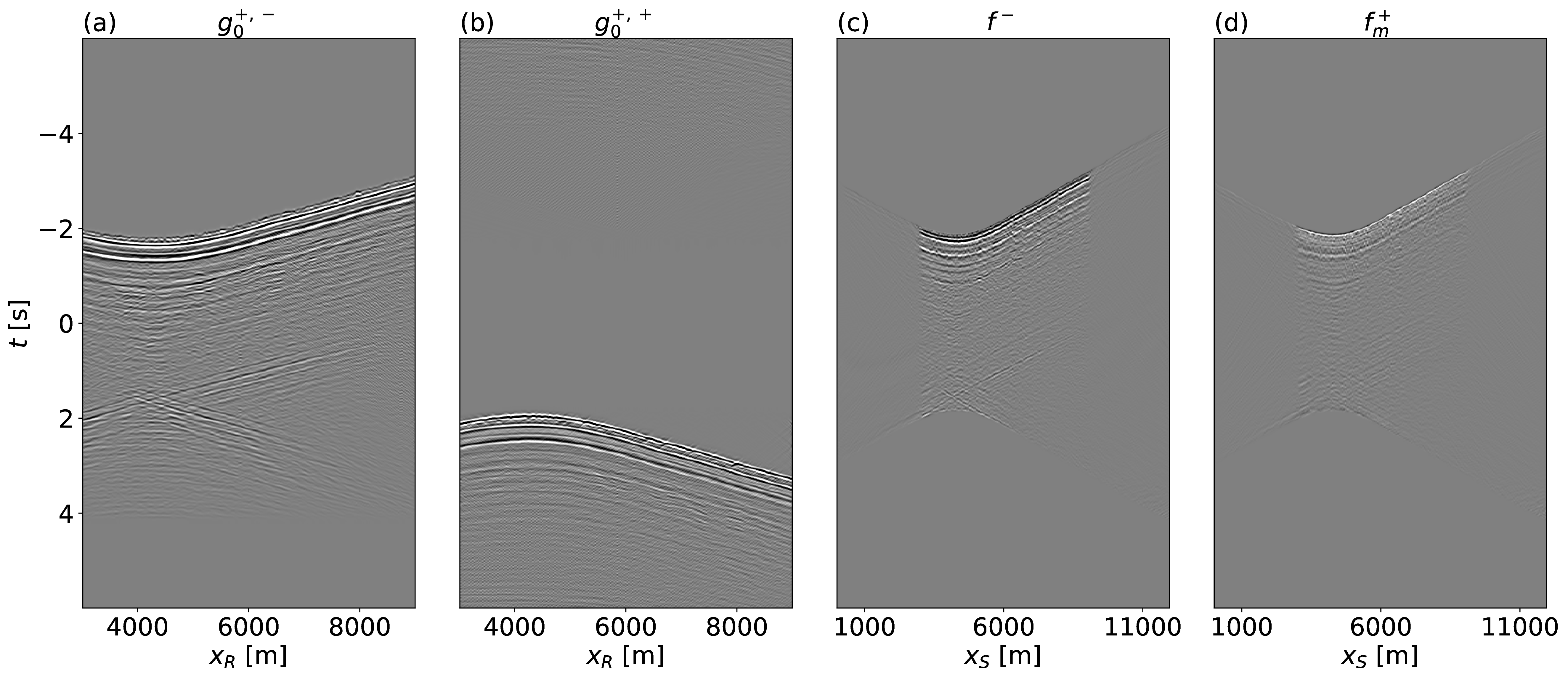}
\caption{(a) Initial up-going subsurface wavefield; (b) initial down-going subsurface wavefield; (c) final up-going focusing function; (d) final down-going focusing function for the Volve field data.}
\label{fig:in_out_put_volve}
\end{figure*}

In this dataset, as horizontal apertures of the sources and receivers differ, the initial subsurface wavefields and the focusing functions exhibit distinct wave propagation paths. Specifically, the subsurface waves propagate from the focal point to the mirrored receivers, while the focusing functions originate from the sources to the focal point. As a result, the structure of the initial subsurface wavefields and focusing functions varies significantly.
To illustrate this clearly, Figure~\ref{fig:in_out_put_volve} shows the initial subsurface wavefields and the corresponding focusing functions for a selected training point, indicated by the red triangle in Figure~\ref{fig:geometry_volve}a. Both the vertical and lateral positioning differences contribute to the increased complexity in the mapping between initial subsurface wavefields and focusing functions. The mismatch between the depth levels at which subsurface wavefields and focusing functions are retrieved already adds complexity when compared to the standard Marchenko method. The additional effect of lateral non-coincidence further complicates the learning task. 
In this context, embedding the spatial position of imaging points into the network becomes beneficial, as it provides crucial contextual information that helps improve the training and generalization performance.
Since the size of the focusing functions for the Volve field data is larger than the synthetic example, the batch size is reduced from 16 to 8, while other parameters in the network remain unchanged. 

The training process is conducted for 250, 500, and 3000 epochs for the three tests, respectively, as their training and validation convergence curves shown in Figure \ref{fig:geometry_volve}b. To assess the impact of positional embedding, Figure~\ref{fig:geometry_volve}c compares the convergence behavior for the 15\% data case, with and without positional information. The results demonstrate that embedding positional information leads to more stable and efficient training.

\begin{figure*}[!t]
\centering
\includegraphics[width=1\textwidth]{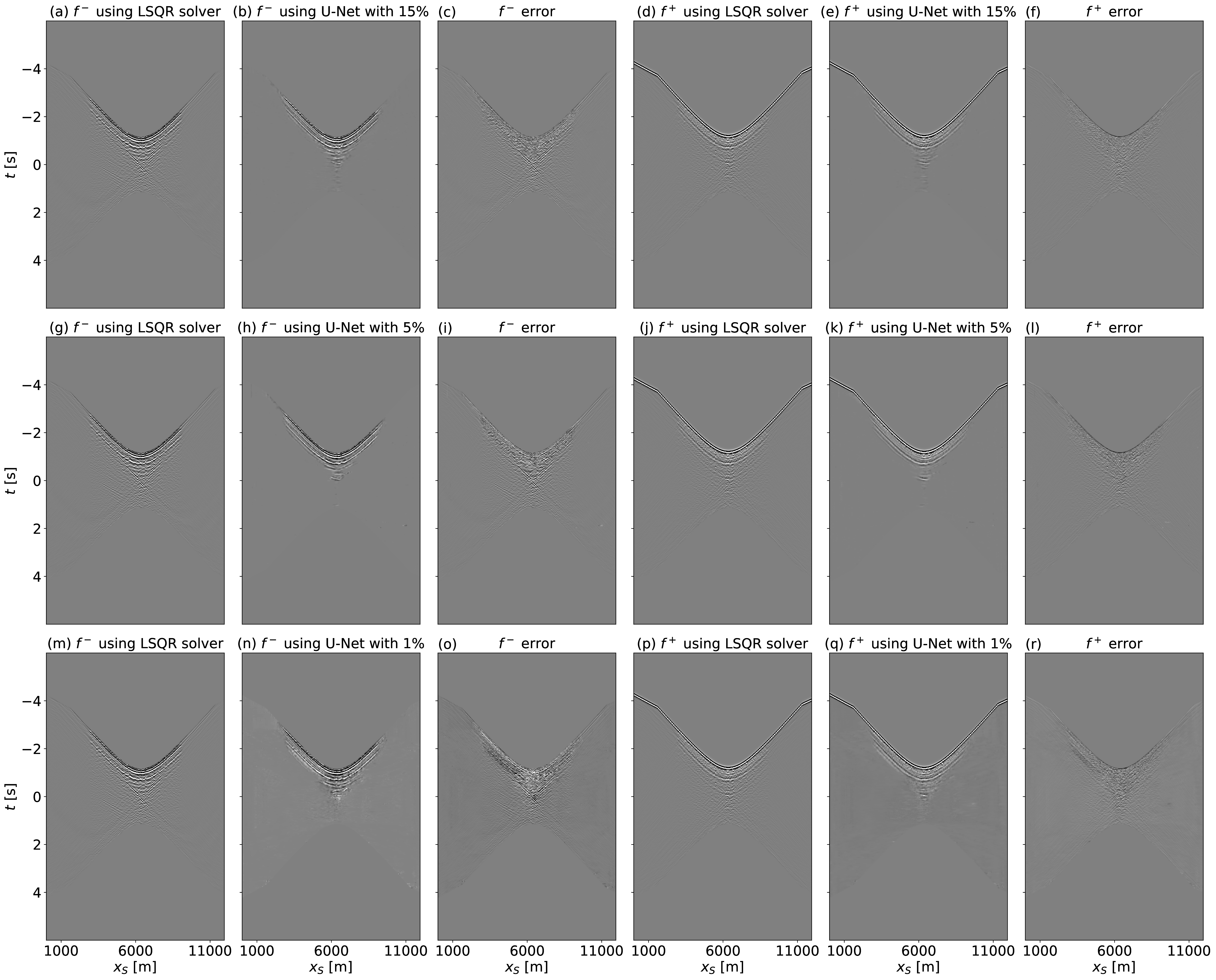}
\caption{Up-going focusing functions obtained using 15\%, 5\%, and 1\% of training and validation data, computed by the LSQR solver (a, g, m) and U-Net (b, h, n), with corresponding differences shown in (c, i, o). Down-going focusing functions for the same cases, computed by the LSQR solver (d, j, p) and U-Net (e, k, q), with corresponding differences shown in (f, l, r).}
\label{fig:f1_inv_volve}
\end{figure*}

The performance of the trained network is evaluated using the predicted results for a selected imaging point marked by the blue star in Figure \ref{fig:geometry_volve}a. The predicted up- and down-going focusing functions, along with the corresponding reference solutions computed using the LSQR solver and their differences, are shown in Figure~\ref{fig:f1_inv_volve}, as results obtained using 15\%, 5\%, and 1\% of training and validation data are presented in the first, second, and third rows, respectively. We can notice how the predicted focusing functions closely resemble their expected counterpart for the main high-amplitude events, while lower-amplitude events are more challenging to capture. However, the overall structure of the predictions remains consistent with the reference focusing functions computed via the LSQR solver.

\begin{figure*}[!t]
\centering
\includegraphics[width=1\textwidth]{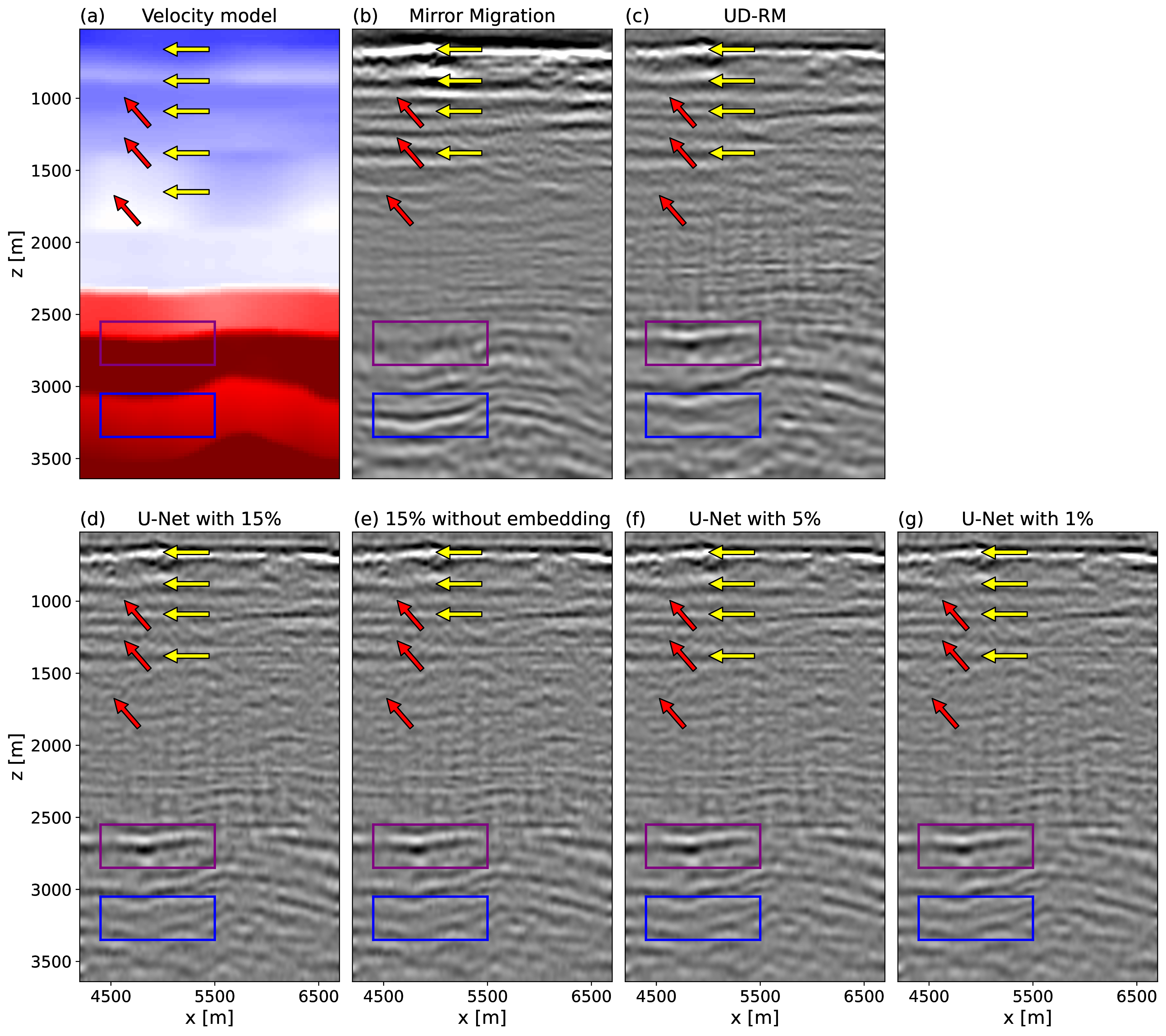}
\caption{(a) Imaging area; (b) mirror migration result; UD-RM imaging results using (c) the LSQR solver and the U-Net trained with (d) 15\%, (e) 15\% without positional embedding, (f) 5\%, and (g) 1\% of the training and validation data.}
\label{fig:Image_volve}
\end{figure*}

Imaging is then performed over the target area shown in Figure~\ref{fig:Image_volve}a, using subsurface wavefields computed by applying equation~\ref{eq:UDRM_matrix} to focusing functions estimated via both the LSQR solver and the U-Net. For comparison, mirror migration is also conducted using the same input wavefields as the UD-RM method (Figure~\ref{fig:Image_volve}b). The mirror migration image contains most of the real reflectors (indicated by yellow arrows); however, it also exhibits prominent artifacts, including ghost reflectors caused by the surface-related multiples (red arrows) and internal multiples (highlighted by the blue box).
In contrast, the images generated from UD-RM wavefields, using both the LSQR solver (Figure~\ref{fig:Image_volve}c) and the U-Net (Figures~\ref{fig:Image_volve}d–\ref{fig:Image_volve}g), successfully suppress most of these artifacts, as indicated by the red arrows and in the blue box. Moreover, these images show better alignment with the reference velocity model (Figure~\ref{fig:Image_volve}a) and improved reflector continuity, particularly in the region highlighted by the purple square.
However, when compared to the LSQR-based result (Figure~\ref{fig:Image_volve}c), the U-Net-generated images exhibit some degradation, especially in the same purple-square region in Figures \ref{fig:Image_volve}d to \ref{fig:Image_volve}g, reflecting the trade-off between computational efficiency and reconstruction fidelity. 
While the loss curves in Figure~\ref{fig:geometry_volve}c clearly demonstrate improved convergence when positional embedding is used, this improvement is less evident in the final imaging results. As shown in Figures~\ref{fig:Image_volve}d and \ref{fig:Image_volve}e, both images exhibit comparable quality. Nevertheless, it is important to note that incorporating positional information contributes to more efficient training in this example.

Finally, we assess the computational performance of the proposed method on the Volve field dataset. The conventional LSQR-based UD-RM approach requires approximately 1 minute 16 seconds per imaging point, resulting in a total processing time of approximately 223 hours 53 minutes for 10605 imaging points.
In contrast, the U-Net-based approach with 15\% of training and validation points reduces the total computation time to approximately 45 hours 8 minutes 51 seconds. This includes 33 hours 34 minutes 57 seconds for computing the focusing functions at 15\% of the imaging points, 11 hours 26 minutes 45 seconds for training the network, and 6 minutes 9 seconds for inference, which achieves approximately a 5-fold reduction in total runtime.
With 5\% of imaging points used for training and validation, the total computation time further decreases to approximately 19 hours 47 minutes 50 seconds, which includes 11 hours 11 minutes 39 seconds for focusing function computation, 8 hours 29 minutes 11 seconds for training, and 7 minutes for inference. This achieves approximately an 11-fold speed-up.
For the test with only 1\% of imaging points for training and validation, the total computational time is reduced to approximately 12 hours 26 minutes 56 seconds. This includes 2 hours 14 minutes 20 seconds for focusing function computation, 10 hours 5 minutes 20 seconds for training, and 7 minutes 16 seconds for inference, yielding an 18-fold reduction in computational time.
These results, consistent with those observed in the synthetic examples, confirm that the proposed method can significantly accelerate the UD-RM workflow while preserving acceptable imaging quality. 

\section*{Conclusions}
In this work, we introduced a self-supervised deep learning framework to accelerate the UD-RM imaging method where the up- and down-going focusing functions are predicted using a U-Net architecture. The proposed approach significantly reduces the computational cost associated with traditional iterative solvers such as LSQR and FISTA, while maintaining imaging quality comparable to the conventional UD-RM method. Evaluations on both synthetic and field datasets demonstrate that the U-Net-based method achieves substantial speed-ups without compromising the ability to suppress internal and surface-related multiples. Even in challenging settings with limited training data and sparse receiver coverage, the method produces images that closely align with those obtained using conventional solvers, capturing key reflectors and reducing artifacts.

The effectiveness of the proposed method comes not only from the architectural capabilities of the U-Net network, but also from the incorporation of physical constraints, such as a traveltime-based windowing operator, embedding information of the imaging point positions, and the use of a normalized MSE loss function that balances the amplitude differences between focusing function components. By relying solely on data from the target imaging area, the framework avoids the need for external training datasets and remains flexible across different acquisition geometries. Overall, this work presents a practical and efficient solution for accelerating UD-RM imaging, transforming it from a computationally intensive approach into an efficient and accessible tool for accurate, large-scale seismic imaging applications.

\section{Acknowledgments}
The authors thank KAUST for providing the computational resources and research support that made this work possible. We also acknowledge Equinor and partners for the release of the Volve dataset. Additionally, we express our gratitude to the DeepWave sponsors for supporting this research. 

\bibliographystyle{seg}  
\bibliography{example}

\end{document}